\documentclass[12pt,preprint]{aastex}

\usepackage{unites2e}

\shorttitle{TeV Gamma Rays from 1ES1959}
\shortauthors{Holder et al.}

\received{2002 November 15}
\begin{document}


\title{Detection of TeV Gamma Rays from the BL Lac Object 1ES1959+650 with the Whipple 10m Telescope}


\author{J.~Holder\altaffilmark{1}, 
I.H.~Bond\altaffilmark{1}, 
P.J.~Boyle\altaffilmark{2},
S.M.~Bradbury\altaffilmark{1},
J.H.~Buckley\altaffilmark{3}, 
D.A.~Carter-Lewis\altaffilmark{4},  
W.~Cui\altaffilmark{5},  
C.~Dowdall\altaffilmark{6},
C.~Duke\altaffilmark{7}, 
I.~de la Calle Perez\altaffilmark{1}, 
A.~Falcone\altaffilmark{5},
D.J.~Fegan\altaffilmark{8},
S.J.~Fegan\altaffilmark{9}\altaffilmark{,10}, 
J.P.~Finley\altaffilmark{5},
L.~Fortson\altaffilmark{2},
J.A.~Gaidos\altaffilmark{5}, 
K.~Gibbs\altaffilmark{9},
S.~Gammell\altaffilmark{8},
J.~Hall\altaffilmark{11},
T.A.~Hall\altaffilmark{12},
A.M.~Hillas\altaffilmark{1}, 
D.~Horan\altaffilmark{9},  
M.~Jordan\altaffilmark{3}, 
M.~Kertzman\altaffilmark{13}, 
D.~Kieda\altaffilmark{11},
J.~Kildea\altaffilmark{8}, 
J.~Knapp\altaffilmark{1},
K.~Kosack\altaffilmark{3},  
H.~Krawczynski\altaffilmark{3},
F.~Krennrich\altaffilmark{4}, 
S.~LeBohec\altaffilmark{4},
E.T.~Linton\altaffilmark{2},
J.~Lloyd-Evans\altaffilmark{1},
P.~Moriarty\altaffilmark{6},  
D.~M\"uller\altaffilmark{2},
T.N.~Nagai\altaffilmark{11},
R.~Ong\altaffilmark{14},   
M.~Page\altaffilmark{7},
R.~Pallassini\altaffilmark{1},
D.~Petry\altaffilmark{4,16},
B.~Power-Mooney\altaffilmark{8},
J.~Quinn\altaffilmark{8},
P.~Rebillot\altaffilmark{3}, 
P.T.~Reynolds\altaffilmark{15},           
H.J.~Rose\altaffilmark{1}, 
M.~Schroedter\altaffilmark{9}\altaffilmark{,10},
G.H.~Sembroski\altaffilmark{5},  
S.P.~Swordy\altaffilmark{2}, 
V.V.~Vassiliev\altaffilmark{11},
S.P.~Wakely\altaffilmark{2},
G.~Walker\altaffilmark{11},  
T.C.~Weekes\altaffilmark{9} }


\altaffiltext{1}{Department of Physics, University of Leeds,
Leeds, LS2 9JT, Yorkshire, England, UK}

\altaffiltext{2}{Enrico Fermi Institute, University of Chicago,  Chicago, IL 60637}

\altaffiltext{3}{Department of Physics, Washington University, St.~Louis,
MO 63130}

\altaffiltext{4}{Department of Physics and Astronomy, Iowa State
University, Ames, IA 50011}

\altaffiltext{5}{Department of Physics, Purdue University, West
Lafayette, IN 47907}

\altaffiltext{6}{School of Science, Galway-Mayo Institute of Technology,
Galway, Ireland}

\altaffiltext{7}{Physics Department, Grinnell College, Grinnell, IA 50112}

\altaffiltext{8}{Physics Department, National University of Ireland,
Belfield, Dublin 4, Ireland}

\altaffiltext{9}{ Fred Lawrence Whipple Observatory, Harvard-Smithsonian
CfA, Amado, AZ 85645}

\altaffiltext{10}{Department of Physics, University of Arizona, Tucson, AZ 85721}

\altaffiltext{11}{High Energy Astrophysics Institute, University of Utah,
Salt Lake City, UT 84112}

\altaffiltext{12}{Department of
Physics and Astronomy, University of Arkansas, Little Rock, AR 72204}

\altaffiltext{13}{Physics Department, De Pauw University, Greencastle, 
                   IN, 46135}

\altaffiltext{14}{Department of Physics, University of California, Los Angeles, CA 90095}

\altaffiltext{15}{Department of Physics, Cork Institute of Technology, Cork, Ireland}

\altaffiltext{16}{NASA/GSFC, Code 661, Greenbelt, MD 20771}


\begin{abstract}
We present the first strong detection of very high energy $\gamma$-rays from
the close ($z=0.048$) X-ray selected BL Lacertae object
1ES1959+650. Observations were made with the Whipple 10m telescope on
Mt. Hopkins, Arizona, using the atmospheric Cherenkov imaging technique. The
flux between May and July 2002 was highly variable, with a mean of
$0.64\pm0.03$ times the steady flux from the Crab Nebula and reaching a maximum of five Crab, with variability on timescales as short as seven hours. 

\end{abstract}




\section{Introduction}

Blazars, consisting of flat spectrum radio quasars (FSRQs) and BL Lacertae
objects (BL Lacs), are a relatively rare type of active galaxy 
characterized by extremely variable, non-thermal spectral energy
distributions (SEDs). The emission extends from radio to $\gamma$-ray
frequencies and is believed to be produced by a highly relativistic plasma jet
closely orientated with the line of sight to the galaxy \citep{Urry95,
Blandford78}. In a $\nu F_{\nu}$ representation, the SEDs display two
peaks. The lower energy peak is usually attributed to synchrotron radiation
from relativistic electrons, while the higher energy peak is thought to be due
to inverse Compton scattering by these same electrons. The photons involved in
the scattering process may be either synchrotron photons produced in the jet,
in the synchrotron self-Compton (SSC) model, or some external population, such
as photons emitted from an accretion disc \citep{Dermer92} or reflected from
emission line clouds \citep{Sikora94}. Alternative explanations for the high
energy component invoke pair cascades triggered via pion and pair
photoproduction from high energy protons in the jet \citep{Mannheim93, Mannheim98}, or
proton synchrotron radiation \citep{Aharonian00, Mucke01}.

The EGRET detector on board the CGRO detected in excess of 66 blazars
\citep{Hartman99} above $100\U{MeV}$; however, only a few have been detected
at higher energies ($>~300\U{GeV}$) by ground based atmospheric Cherenkov
telescopes. Markarian~421 \citep{Punch92} and Markarian~501 \citep{Quinn96}
were the first extragalactic TeV sources to be detected and the behaviour of
these two nearby ($z=0.031$ and $z=0.034$ respectively) objects has been closely
monitored and studied. Flux variability over two orders of magnitude has been
observed, with doubling timescales as short as $\sim15\U{minutes}$
\citep{Gaidos96}, placing strong constraints on the size of the emission
region. The $\gamma$-ray flux is correlated with the X-ray flux
\citep{Buckley96, Catanese97a, Krawczynski97} and, in the case of Markarian 421,
with the spectral power law index in the $\gamma$-ray region
\citep{Krennrich02,Aharonian02a}. Unconfirmed detections of 1ES 2344+514 \citep{Catanese98}
and PKS 2155-304 \citep{Chadwick99} have also been reported and, most
recently, the detection of very weak emission from H1426+428 has been
confirmed by three groups \citep{Horan02, Aharonian02b, Djannati02}. This last
object is of particular interest since it is relatively distant ($z=0.129$) and
hence the observed flux is expected to be strongly attenuated by absorption
via pair production on the extragalactic infra-red background light (EBL)
\citep{Gould67, Stecker92}.

All of the TeV blazars can be classified as high frequency peak BL 
Lacs (HBLs) \citep{Padovani95} on the basis of the location of the lower energy
peak in their SEDs, and predictions of TeV emission from other nearby sources
of this type have driven observations by atmospheric Cherenkov
detectors. 1ES1959+650 ($z=0.048$) was first suggested as a good TeV candidate
of this type by \citet{Stecker96}, who used simple scaling arguments to
predict it as the third strongest TeV blazar, after
Markarian~421 and Markarian~501, with a flux prediction of
$1.9\times10^{-11}\UU{cm}{-2}\UU{s}{-1}$ above $300\U{GeV}$. More recently \citet{Costamante02}
have predicted fluxes above $300\U{GeV}$ of
$7.5\times10^{-11}\UU{cm}{-2}\UU{s}{-1}$, based on a simple phenomenological
parameterization of the SED adapted from \citet{Fossati98}, and
$0.03\times10^{-11}\UU{cm}{-2}\UU{s}{-1}$ using a homogeneous, one-zone
synchrotron self-Compton model \citep{Ghisellini02}. 

In fact, 1ES1959+650 is not a particularly extreme example of an
HBL. Measurements with BeppoSAX in 1997 \citep{Beckmann02} placed the lower
energy SED peak at $10^{15}\U{Hz}$ ($4\U{eV}$), as compared to typically
$\sim1\U{keV}$ for Markarian~421 \citep{Maraschi99} and $\sim100\U{keV}$ for
Markarian~501 during its 1997 flare state \citep{Pian98}, although for both of
these sources the position of the peak is known to vary widely depending upon
the flux level. \citet{Giebels02} report on X-ray observations of 1ES1959+650 obtained with the
USA(1-16\U{keV}) and RXTE (2-16\U{keV}) missions during 2000, showing
threefold increases in the X-ray flux on a timescale of a few days. The flux
increase is correlated with spectral hardening, indicative of a shift of the
lower energy peak of the SED towards higher frequencies. The source is also
unusual among HBLs for its strong rapid optical variability. Observations by
\citet{Villata00} showed rapid flickering, including a decrease of
0.28~magnitudes in four days.

Prior to May 2002, only tentative evidence had been presented for TeV emission from 1ES1959+650. A detection with a
statistical significance of $3.9\U{\sigma}$ was reported by \citet{Nishiyama99}
based on $57\U{hours}$ of observations with the Utah Seven Telescope
Array. More recently, \citet{Konopelko02} reported a preliminary detection at
$\sim5\U{\sigma}$ for the HEGRA Cherenkov telescope system. Previous
observations of 1ES1959+650 with the Whipple 10m telescope produced
an upper limit at a flux level of $1.3\times10^{-11}\UU{cm}{-2}\UU{s}{-1}$
above $350\U{GeV}$ \citep{Catanese97b}.

Observations of 1ES1959+650 during May-July 2002 with the Whipple 10m
telescope resulted in a clear detection of TeV $\gamma$-ray emission from this
object \citep{IAUC} which was rapidly confirmed by the HEGRA experiment
\citep{IAUC2}. We report here the results of the Whipple 10m observations.

\section{Observations and Data Analysis}

The configuration of the Whipple 10m $\gamma$-ray telescope during these
observations is described in detail in \citet{Finley01}. Briefly, the
telescope consists of a $10\U{m}$ reflector and a 490 pixel photomultiplier
tube (PMT) camera. For the purposes of this analysis only the high resolution
($0.12^{\circ}$ spacing) central 379 PMT pixels have been used. Cherenkov
images are recorded and parameterized according to \citet{Hillas85} and then
$\gamma$-ray like images are selected using the ``supercuts'' criteria
\citep{Reynolds93} optimized on recent data from the Crab Nebula, the standard
candle of TeV $\gamma$-ray astronomy \citep{Weekes89}. Following a realignment
of the optical system in February 2002, observations of the Crab Nebula showed
that the telescope was at its most sensitive since the installation of the
current camera, with one hour of Crab observations producing a $6\sigma$
detection; however, between February and July 2002 a decrease in the
relative efficiency of the telescope of $\sim30\%$ was measured by examining
the response to the cosmic ray background. The cause of the effect is still
under investigation - one explanation may be that it is due to increased
atmospheric absorption caused by large forest fires in the region during this 
period. 

\begin{figure}
\plotone{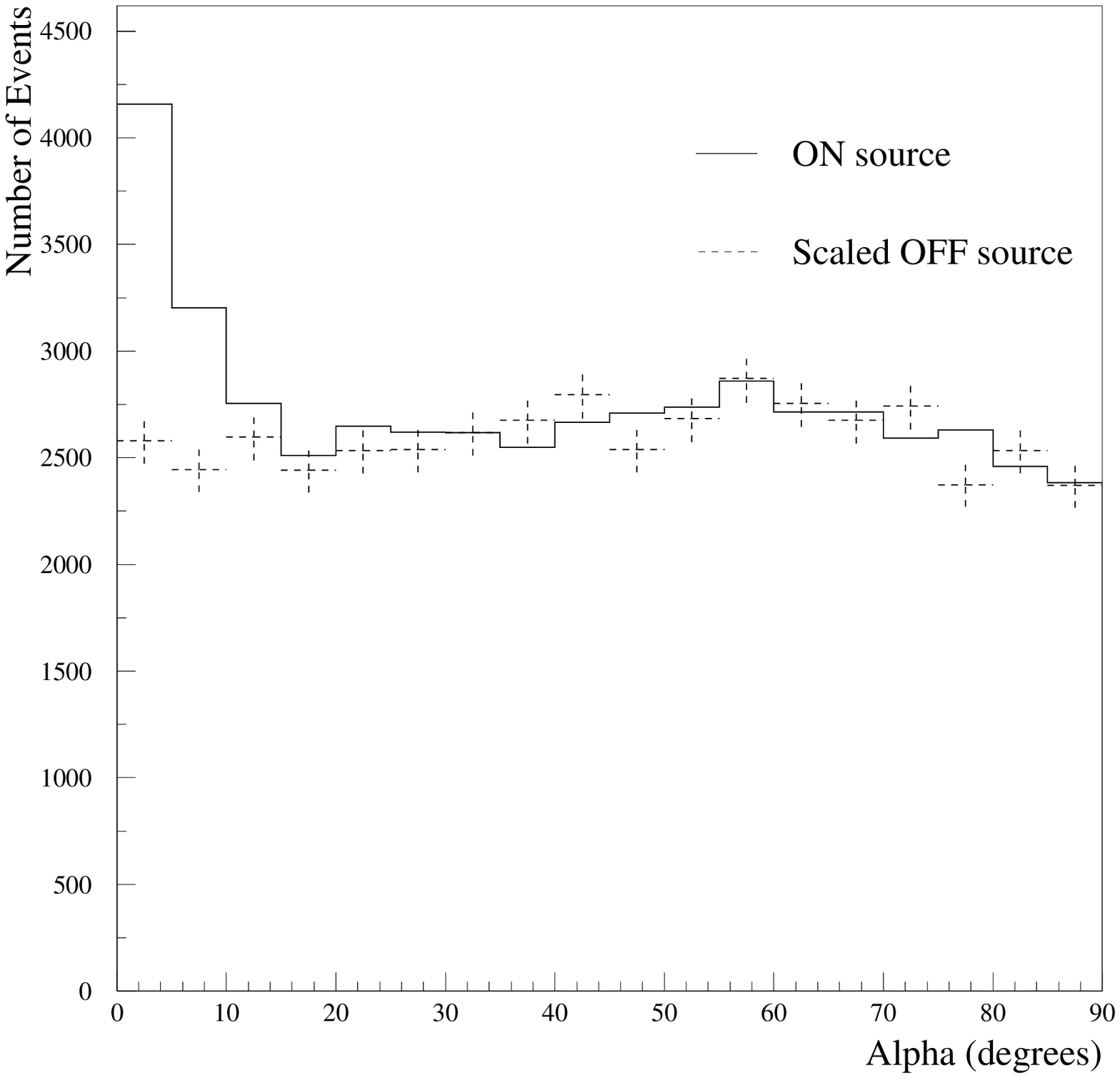}
\caption{Distribution of $\textit{alpha}$ for all on and off-source events
(the off-source distribution has been scaled so as to match the number of
events in the $\alpha>30^{\circ}$ region). $\textit{Alpha}$ is defined as the
angle between the major axis of an elliptical $\gamma$-ray image and the
source location in the camera. \label{alpha}}
\end{figure}

Observations of 1ES1959+650 were made during moonless periods between May
$16^{\mathrm{th}}$ and July $8^{\mathrm{th}}$ 2002. The total dataset consists
of $39.3\U{hours}$ of on-source data, together with $7.6\U{hours}$ of
off-source data for background comparison. For observations from Mt. Hopkins
(latitude $31^{\circ}57.6'$N) 1ES1959+650 culminates at a zenith angle
($\theta_z$) of $33.5^{\circ}$ and so the data were necessarily taken at large
$\theta_z$, between $53.5^{\circ}$ and $33.5^{\circ}$. After accounting for the
 zenith angle and reduced telescope efficiency, we estimate the energy 
threshold ($E_{thresh}$) for the majority of the observations to be $\sim600\U{GeV}$
. $E_{thresh}$ is defined here as the energy of the peak differential 
$\gamma$-ray flux for a source with the same spectrum as the Crab Nebula. 
Figure~\ref{alpha}
shows the distribution of the Cherenkov image orientation angle \textit{alpha}
for the full on-source dataset, together with the distribution for the
off-source dataset scaled such that the number of events in the region
$\alpha>30^{\circ}$ is the same. The average rate of the excess in the
on-source $\gamma$-ray region at $\alpha<15^{\circ}$ is $1.08\pm0.05
\U{\gamma}\UU{min}{-1}$, corresponding to a detection of greater than $20\U{\sigma}$.

\section{Flux Variability}
\begin{figure}
\plotone{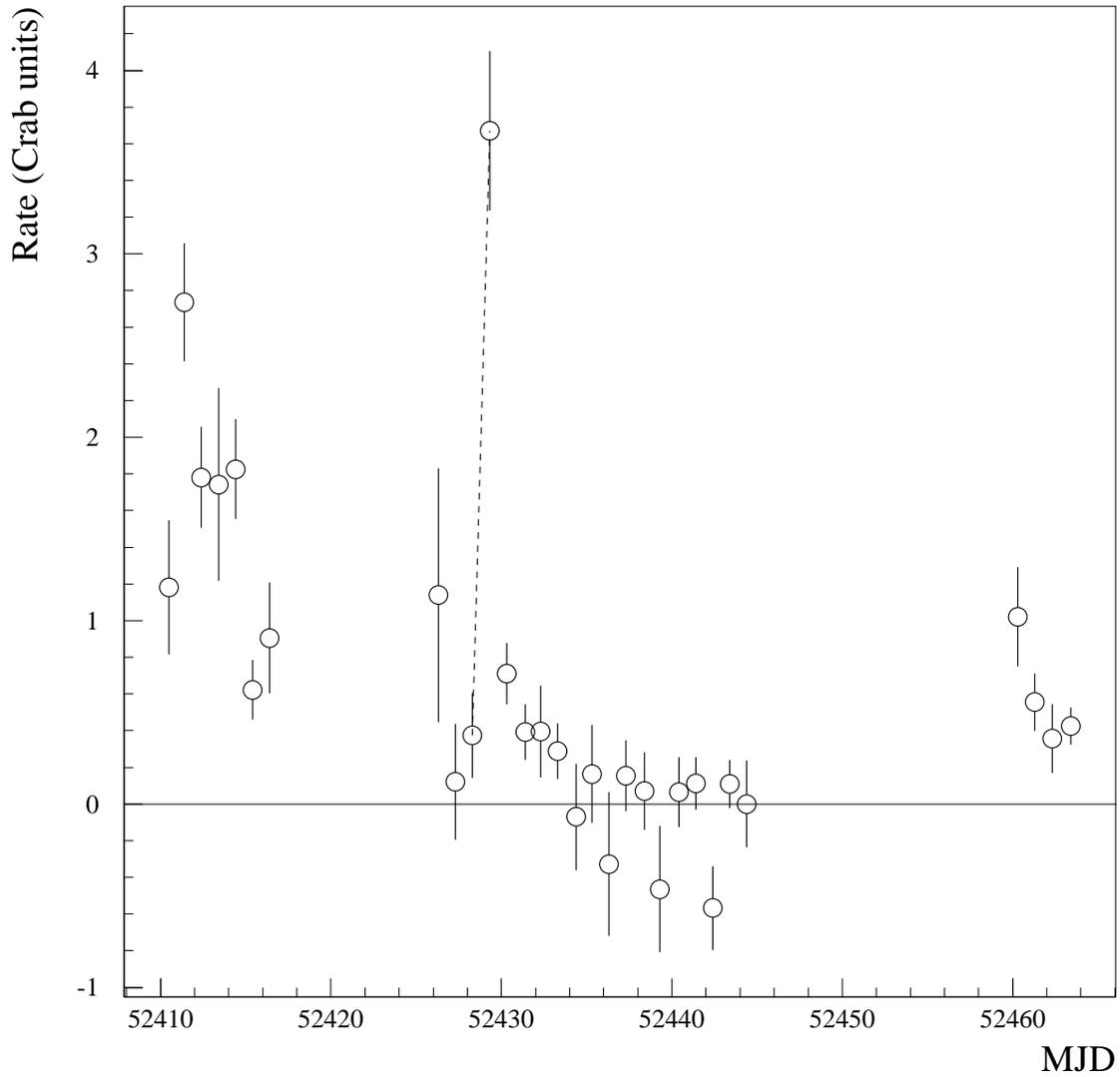}
\caption{The daily average $\gamma$-ray rates for 1ES1959+650 during
2002. The rates have been corrected for zenith
angle of observation and relative telescope efficiency as described in the
text. The dashed line indicates the most rapid rate change, corresponding to a doubling time of $7\U{hours}$. \label{daily}}
\end{figure}

Figure~\ref{daily} shows the daily averaged rates for
1ES1959+650 for all of the observations. The rates are expressed in multiples of
the steady Crab Nebula flux and have been corrected in order to account for
the varying zenith angle and changes in telescope efficiency according to
the procedure of \citet{Lebohec02}. This has been tested using Crab Nebula
data taken over a wide range of zenith angles and atmospheric conditions. The correction factor varies from run to run and has a mean value for this dataset of $2.0\pm20\%$. Strong
night-to-night variability is evident; the largest change in rate, between MJD
52428 and MJD 52429, corresponds to a doubling time of $7\U{hours}$ - shorter
than has ever been observed in other wavebands for this source. The mean flux
over all observations was $0.64\pm0.03$ Crab units.

\begin{figure}
\plotone{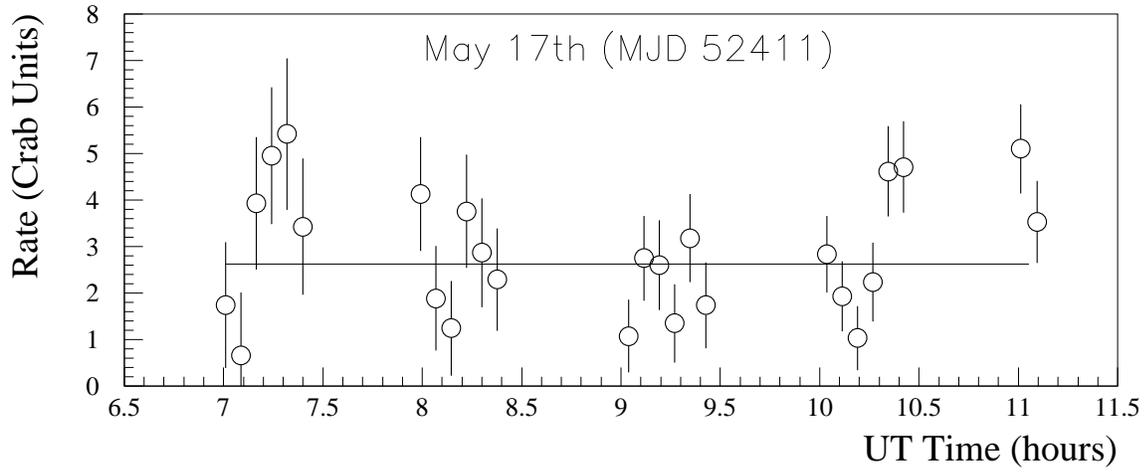}
\plotone{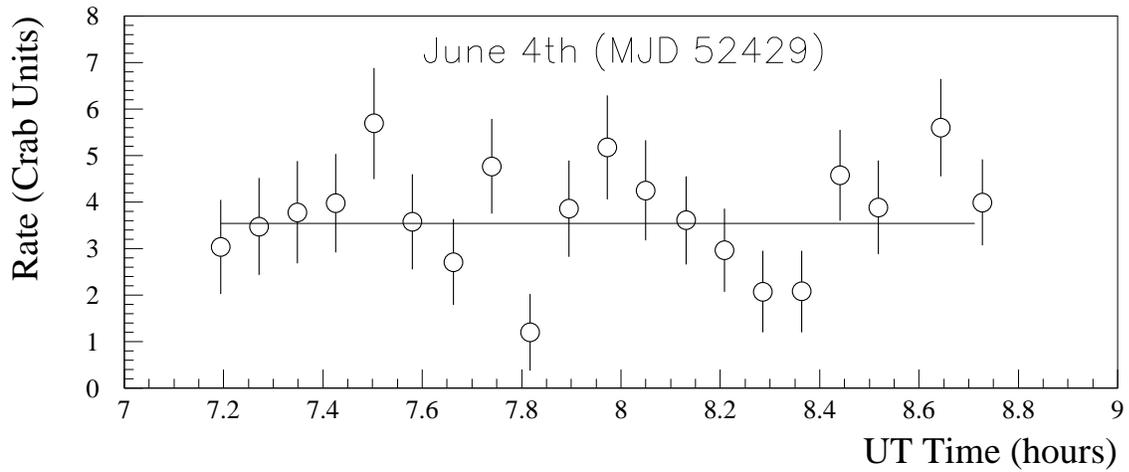}
\caption{The $\gamma$-ray rates for the two nights showing the most
activity ($5\U{minute}$ binning). The rates have been corrected for zenith
angle of observation and relative telescope efficiency as described in the
text. \label{nightly}}
\end{figure}

Figure~\ref{nightly} shows the rate in $5\U{minute}$ bins for
two nights, May $17^{\mathrm{th}}$ and June $4^{\mathrm{th}}$, during which
the source was most active. The statistical evidence for variability within
these nights is given by the $\chi^{2}$ probabilities of constant emission
= 1\% and 8\% respectively. We conclude that there is no strong evidence for
flux variability on this timescale.
 
\section{Discussion}

The detection of TeV $\gamma$-ray emission from 1ES1959+650 adds another
member to the class of TeV blazars, all of which are close BL Lac objects
having a low bolometric luminosity and the peaks in their SEDs at high
frequency. The rapid flux variability, and the fact that 1ES1959+650 has not
been detected during previous observations, indicates that the source was in
an unusual flaring state during these observations. The flux level was at
times orders of magnitude above the most recent model predictions. Throughout the
period of the Whipple observations measurements in the $2-12\U{keV}$ region by
the All-Sky Monitor (ASM) on board RXTE have shown 1ES1959+650 to be active
and variable, with daily average fluxes reaching $\sim20\U{mCrab}$ in May and
July. Target of opportunity observations with the RXTE small field of view
instruments were triggered following the $\gamma$-ray detection and will be
reported on elsewhere (Krawczynski, private communication). The rapid flux
variability observed at TeV energies implies a small emission region in the
jet with a high Doppler factor \citep{Mattox93,Madejski96,Buckley96}. The
contemporaneous X-ray and TeV $\gamma$-ray data will allow us to constrain the jet
parameters when modelling the emission processes.

Analysis of the Whipple data is ongoing, but attempts to reconstruct the
source spectrum have been hampered by the effects of the decreased telescope efficiency during the observation period. The HEGRA collaboration measure a
rather steep spectrum (spectral index $\alpha=3.2\pm0.3$) for observations
prior to 2002, while the spectrum during the flaring period exhibits
pronounced curvature and deviates significantly from the spectrum seen during
the quiescent state \citep{Horns02}. The majority
of models for the EBL lead to predictions of a cut-off in the $\gamma$-ray
region beginning below $\sim10\U{TeV}$ for a source at $z=0.048$
 (\citet{Primack01}; but see \citet{Vassiliev00} for a discussion of 
an EBL model which does not produce a distinct feature in the observed
spectrum). 
Deviations from a pure power law have now
been resolved in the spectra of both Markarian~421 \citep{Krennrich01, Piron01,
  Aharonian02a} and
Markarian~501 \citep{Samuelson98, Aharonian99, Djannati99} and the spectrum of
the most distant TeV blazar, H1426+428, is measured to be very steep
($\alpha=3.50\pm0.35\mathrm{(stat)}\pm0.05\mathrm{(syst)}$) \citep{Petry02},  
but it is not yet clear whether these features are due to
absorption on the EBL or are intrinsic to the sources. Clearly, further
observations and spectral analysis of 1ES1959+650 may help to resolve this
question.

\acknowledgments
The VERITAS Collaboration is supported by the U.S. Dept. of Energy, N.S.F.,
the Smithsonian Institution, PPARC (U.K.) and Enterprise Ireland.

\clearpage


\clearpage 

\clearpage 


\end{document}